\documentclass[%
 reprint,
 amsmath,amssymb,
 aps,
]{revtex4-1}

\usepackage{caption}
\usepackage{colordvi}
\usepackage{amsmath}
\usepackage{graphicx}

\usepackage{graphicx}
\usepackage{dcolumn}
\usepackage{bm}
\usepackage{enumerate}
\usepackage{multirow}
\usepackage{cancel}
\usepackage[table]{xcolor}
\usepackage{tabularx}
\usepackage[para]{threeparttable}
\usepackage{dcolumn}
\usepackage{amssymb}
\usepackage{afterpage}
\usepackage{nicefrac}
\usepackage{mhchem}
\usepackage{physics}
 
\newcommand{\Lower}[1]{\smash{\lower 1.5ex \hbox{#1}}}

\usepackage{ulem}
\newcommand{\down}{\sout{$\downarrow$}}
\newcommand{\up}{\sout{$\uparrow$}}
\newcommand{\zero}{\sout{\phantom{$\downarrow \negthickspace \uparrow$}}}
\newcommand{\double}{\sout{$\downarrow \negthickspace \uparrow$}}
\newcommand{\ib}{\bar{i}}
\newcommand{\jb}{\bar{j}}

\usepackage{soul}

\begin{document}

\preprint{APS/123-QED}
\title{
Orbital entanglement and correlation from pCCD-tailored Coupled Cluster wave functions}%
\author{Artur Nowak}
\affiliation{Institute of Physics, Faculty of Physics, Astronomy and Informatics,
Nicolaus Copernicus University in Toru\`n, Grudziadzka 5, 87-100 Torun, Poland
\\}
\vspace{-2cm}
\author{\"Ors Legeza}
\affiliation{%
 Strongly Correlated Systems ``Lend\"ulet" Research Group, Wigner Research Center for Physics, H-1525 Budapest, Hungary\\}
\author{Katharina Boguslawski}
\email{k.boguslawski@fizyka.umk.pl}
\affiliation{Institute of Physics, Faculty of Physics, Astronomy and Informatics,
Nicolaus Copernicus University in Toru\`n, Grudziadzka 5, 87-100 Torun, Poland
\\}
\vspace{-2cm}
 \author{}
\affiliation{%
 \\}


\date{\today}

\begin{abstract}
Wave functions based on electron-pair states provide inexpensive and reliable models to describe quantum many-body problems containing strongly-correlated electrons, given that broken-pair states have been appropriately accounted for by, for instance, \textit{a posteriori} corrections.
In this article, we analyse the performance of electron-pair methods in predicting orbital-based correlation spectra.
We focus on the (orbital-optimized) pair-coupled cluster Doubles (pCCD) ansatz with a linearized coupled-cluster~(LCC) correction.
Specifically, we scrutinize how orbital-based entanglement and correlation measures can be determined from a pCCD-tailored CC wave function.
Furthermore, we employ the single-orbital entropy, the orbital-pair mutual information, and the eigenvalue spectra of the two-orbital reduced density matrices to benchmark the performance of the LCC correction for the one-dimensional Hubbard model with periodic boundary condition as well as the \ce{N2} and \ce{F2} molecules against DMRG reference calculations.
Our study indicates that pCCD-LCC accurately reproduces the orbital-pair correlation patterns in the weak-correlation limit and for molecules close to their equilibrium structure.
Hence, we can conclude that pCCD-LCC predicts reliable wave functions in this regime.
In the strong-correlation limit and for molecules with stretched bonds, the LCC correction, generally, overestimates orbital-pair correlations.

\end{abstract}

\pacs{Valid PACS appear here}
\maketitle

\section{Introduction}
The compromise between the computational cost of quantum-many-body methods and their accuracy and reliability is a central issue in quantum physics and chemistry.
Specifically in electronic structure theory, we require methods that allow us to efficiently describe the correlated motion of electrons.
Difficulties originate from the different contributions that govern the correlated motion of electrons, commonly referred to as strong and weak correlation.
Conventional methods that can model strong electron correlation are based on a multireference ansatz. Examples are the complete active space self consistent field (CASSCF) method \cite{van_besien_2006, theo_casscf_paulovic}, multireference coupled cluster (MRCC) approaches \cite{bogus_mrcc,monika_mrcc, adamowicz-mrcc}, the density matrix renormalization group (DMRG) algorithm \cite{white,white2,white-qc,ors_springer,marti2010b,chanreview,ors_ijqc, hachmann_h50, dmrg_kurashige, dmrg-9,yanai-review, baiardi-reiher, gunst-dmrg, ma-dmrg}, and quantum Monte Carlo (QMC) methods \cite{qmc,qmc-book-chapter}.
Despite their applicability to problems with strongly-correlated electrons, conventional multireference methods (like conventional multiconfigurational SCF theory) typically scale exponentially with the size of the system, which limits their application to small- or medium-sized model systems.
A different group of approaches suitable for strongly-correlated electrons uses non-interacting electron pairs, so-called geminals, to construct the electronic wave function, which is an antisymmetric product of geminals \cite{hurley_1953}.
Some well known geminal-based methods are the antisymmetric product of strongly orthogonal geminals (APSG) \cite{hurley_1953,parks_1958}, the antisymmetric product of interacting geminals (APIG) \cite{coleman_1965,bratoz1965,silver_1969,silver_1970,apig-1,apig-2,surjan-bond-1984,surjan-bond-1985,surjan-bond-1994,surjan-bond-1995,surjan_1999,surjan-bond-2000,surjan2012}, and the antisymmetric product of 1-reference orbital geminals \cite{limacher_2013,oo-ap1rog}, also known as the pair-couled cluster doubles (pCCD) ansatz \cite{tamar-pcc}.
Numerical studies suggest that geminal-based approaches can accurately model systems where strong correlation is important, like the one-dimensional Hubbard model \cite{oo-ap1rog,boguslawski2016} or molecules with stretched bonds \cite{surjan2012,zoboki2013,pawel_jpca_2014,tamar-pcc,frozen-pccd,boguslawski2015,ijqc-eratum,garza2015actinide,kasia-lcc}, even those containing lanthanide \cite{yb2_pawel} or actinide atoms \cite{pawel_pccp2015,garza2015actinide,nowak-pccp-2019}.
Specifically, pCCD is a good approximation to the doubly occupied configuration interaction (DOCI) method \cite{weinhold1967a}, requires, however, only mean-field cost.
The pCCD wave function can be written as a product of geminal creation operators $\phi_{i}^\dagger$ acting on the vacuum state
\begin{equation}\label{eq:ap1rog}
         |\textrm{pCCD} \rangle = \prod_{i=1}^{P}\phi_{i}^\dagger | 0\rangle, 
\end{equation}
with $P$ being the number of electron pairs and
\begin{equation}\label{eq:phi}
\phi_{i}^\dagger =a^\dagger_i a^\dagger_{\bar{i}} + \sum_{a}^{\rm virt}c_{i}^{a}a^\dagger_a a^\dagger_{\bar{a}},
\end{equation}
where the sum runs over all virtual orbitals $a$ and $\{c_i^a\}$ are the geminal expansion coefficients.
In eq.~\eqref{eq:phi}, $i$ ($\bar{i}$) indicates spin-up (spin-down) electrons.
The special form of the pCCD geminal creation operator eq.~\eqref{eq:phi} allows us to rewrite the pCCD wave function in terms of one-particle functions using an exponential ansatz \cite{limacher_2013}
\begin{align}\label{eq:pccd}
|\textrm{pCCD} \rangle  	&= \exp (\sum_{i}^{\rm occ} \sum_{a}^{\rm virt} c_i^a a^\dagger_a a^\dagger_{\bar{a}} a_{\bar{i}} a_i) | \Phi _0\rangle  \nonumber \\
													&= \exp(\hat{T}_\textrm{p})\ket{\Phi_0},
\end{align}
where $|\Phi_0 \rangle$ is some independent-particle wave function (for instance, the Hartree--Fock (HF) determinant) and $\hat{T}_\textrm{p}$ is the electron-pair cluster operator. Note that the cluster operator is restricted to electron pair excitations and the coupled cluster (electron-pair) amplitudes are equivalent to the pCCD geminal coefficients.
Furthermore, the first sum runs over all occupied orbitals, whose number is equivalent to the number of electron pairs.
The exponential form ensures the proper (linear) scaling of the method with the number of electrons (size-extensivity). Size-consistency can be recovered by optimizing the one-particle basis functions \cite{oo-ap1rog, tamar-pcc,ps2-ap1rog,ap1rog-jctc}.

Like other geminal-based methods \cite{pernal2014, surjan2012,jeszenszki2014,ellis2013,frozen-pccd, rassolov2002}, pCCD misses a large fraction of the weak electron correlation effects that are associated with broken-pair states. This deficiency can be depleted by considering \textit{a posteriori} corrections that are typically applied on top of a pCCD reference function. In so-called post-pCCD methods, pCCD is combined with, for instance, single- and multi-reference perturbation theory \cite{piotrus_pt2,AP1roG-PTX}, density functional theory \cite{garza-pccp,garza2015,garza2015actinide}, or (linearized) coupled cluster corrections \cite{frozen-pccd,kasia-lcc}.
While numerous numerical studies on pCCD and post-pCCD exist, which demonstrate their good performance in contrast to conventional electronic structure methods \cite{pawel_jpca_2014,tamar-pcc,frozen-pccd,boguslawski2015,garza2015actinide,kasia-lcc,AP1roG-PTX,eompccderratum,eom-pccd-lccsd,nowak-pccp-2019,pccd_noncovalent}, these studies mostly focus on total or relative energies and energy-derived quantities (counter-examples can be found, for instance, in Refs.~\citenum{oo-ap1rog,pawel_pccp2015,boguslawski2016}).
To the best of our knowledge, an in-depth analysis of electron correlation effects in post-pCCD methods has not been presented, yet.
Thus, the goal of this work is to investigate  the accuracy and reliability of selected post-pCCD methods in describing electron correlation going beyond the rather oversimplified picture of energies and energy-derived quantities.
For that purpose, we will employ concepts of quantum information theory (QIT).
Specifically, QIT will allow us to interpret the approximate electronic wave functions using various measures of orbital entanglement and orbital-pair correlation.
In this work, we will focus on the so-called single-orbital entropy \cite{dmrg-10,roland-runo, ziesche_1995,qit-concepts, schilling-1rdm, vlatko, schilling2014quantum, Ding2020} and the orbital-pair mutual information \cite{dmrg-10,legeza2006,rissler2006,stein-reiher-oe, LUO-qit}.
These orbital entanglement and correlation measures can be applied to dissect electronic wave functions \cite{barcza_11,entanglement_letter,entanglement_bonding_2013,cuo_dmrg,pccp_bonding,boguslawski2015,ijqc-eratum,corinne_2015,zhao2015,boguslawski2017} and thus to assess the quality of electronic structure calculations.
Recently, some of us have analyzed the accuracy and reliability of the pCCD model in describing electron correlation effects using the single-orbital entropy and orbital-pair mutual information \cite{boguslawski2016}. Specifically, the pCCD ansatz leads to significant over-correlation, despite providing reasonable energies.
It remains uncertain whether an \textit{a posteriori} correction is able to cure the deficiencies in electron correlation effects predicted by the pCCD approach.
As post-pCCD method, we will choose the recently presented pCCD-LCC models \cite{kasia-lcc} as they represents a promising alternative to conventional multi-reference electronic structure methods for both electronic ground \cite{kasia-lcc} and excited states \cite{eom-pccd-lccsd,nowak-pccp-2019}, where they statistically outperformed conventional CC methods.
By scrutinizing the orbital entanglement and orbital-pair correlations predicted by pCCD-LCC, we will be able to complement previous numerical studies on the performance and reliability of the LCC corrections.
Additional examples where the application of orbital entanglement and correlation is profoundly insightful are the monitoring of chemical reactions \cite{corinne_2015,zhao2015}, the identification of transition states \cite{corinne_2015}, the analysis of chemical bond orders \cite{entanglement_bonding_2013,pccp_bonding,bonding_qit,boguslawski2015,ijqc-eratum}, and computational protocols aspiring black-box computational setups \cite{entanglement_letter,stein2016,boguslawski2017,autocas,dmrg-10}.

This work is organized as follows. In section~\ref{sec:lccsd}, we will briefly discuss various LCC corrections on top of pCCD.
The calculation of orbital entanglement and correlation measures for a pCCD-LCC wave function is scrutinized in section~\ref{sec:i12}. Section~\ref{sec:i12-lcc} represents the correlation measures in terms of LCC response density matrices. Computational details are showed in section~\ref{sec:details}. Numerical results are presented in section~\ref{sec:results}. Finally, we conclude in section~\ref{sec:conclusions}.

\section{LCC Corrections with a pCCD Reference Function}\label{sec:lccsd}

In pCCD-LCC methods, dynamical correlation is accounted for using a coupled cluster ansatz where the HF reference determinant is substituted by the pCCD reference function,
\begin{equation}
    \ket{\Psi}=  \exp (\hat{T}) \ket{\textrm{pCCD}},
\end{equation} 
where $\hat{T}=\sum_{\nu}t_{\nu}\hat{\tau}_{\nu}$ is some general cluster operator, that is, a sum over excitation operators $\hat{\tau}_\nu$. The electronic energy can be obtained by solving the time-independent Schr\"odinger equation
\begin{align}
    \hat{H} \vert \Psi \rangle & = E \vert \Psi \rangle \nonumber \\
    \hat{H} \exp(\hat{T}) \vert {\rm pCCD} \rangle &= E\exp(\hat{T}) \vert \textrm{pCCD} \rangle.
\end{align}
The above equation can be solved using techniques of single-reference coupled cluster theory.
As we seek a linearized coupled cluster correction, we have to truncate the resulting Baker--Campbell--Hausdorff (BCH) expansion after the second term to arrive at
\begin{equation}\label{eq:lcc}
           (\hat{H} + [\hat{H},\hat{T}]) \vert \textrm{pCCD}\rangle =  E \vert \textrm{pCCD} \rangle.
\end{equation}
In pCCD-LCCD, the cluster operator is chosen as $\hat{T}=\hat{T}^\prime_2$, while pCCD-LCCSD also accounts for single excitations, $\hat{T}=\hat{T_{1} }+ \hat{T}^\prime_2$. The ``$\prime$'' indicates that pair excitations are excluded in the cluster operator as they are already accounted for in pCCD.
To stress the exclusion of electron-pair excitations in the general cluster operator $\hat{T}$, we will employ the notation $\hat{T}^\prime$ throughout this paper.
Using the relation in eq.~\eqref{eq:pccd}, we can rewrite eq.~\eqref{eq:lcc} in terms of a single reference coupled cluster ansatz,
\begin{equation}\label{eq:lccsd/pccd}
           (\hat{H} + [\hat{H},\hat{T}^\prime]) \exp(\hat{T}_{\rm p})\vert \Phi_0 \rangle=  E \exp(\hat{T}_{\rm p})\vert \Phi_0 \rangle,
\end{equation}
where $\hat{T}_\textrm{p}$ is again the electron-pair cluster operator defined in eq.~\eqref{eq:pccd}.
We can further simplify the above equation exploiting that the cluster operators $\hat{T}^\prime$ and $\hat{T}_\textrm{p}$ commute,
\begin{align}\label{eq:lcc+pccd}
            (\hat{H} &+ [\hat{H},\hat{T}^\prime]  + [[\hat{H},\hat{T}^\prime],\hat{T_p}] \nonumber \\
            &+ [\hat{H},\hat{T}_\textrm{p}] + \frac{1}{2} [[\hat{H},\hat{T}_{\rm p}],\hat{T}_{\rm p}]) \vert \Phi_0 \rangle=  E \vert \Phi_0 \rangle,
\end{align}
where we have explicitly included the linear and quadratic terms from the pCCD reference function (the last two terms on the left-hand-side, assuming at most double excitations in $\hat{T}^\prime$) as the exponential ansatz of pCCD is not linearized.
From eq.~\eqref{eq:lcc+pccd}, the working equations for the electronic energy as well as for the singles and (broken-pair)  doubles amplitudes can be derived using projection techniques \cite{kasia-lcc}.
In the following, we will employ well-known tools of conventional CC theory.
Thus, it will be convenient to rewrite eq.~\eqref{eq:lcc+pccd} using an exponential ansatz of the form
\begin{equation}\label{eq:lcc+pccd-exp}
\{ e^{-\hat{T}^\prime -\hat{T}_p} \hat{H}  e^{\hat{T}_p+\hat{T}^\prime} \}_{L^\prime} \ket{\Phi_0} =  E \ket{\Phi_0}.
\end{equation}
Curly brackets indicate that all broken-pair contributions (here, $\hat{T}_1$ and $\hat{T}_2^\prime$ or just $\hat{T}_2^\prime$) appear at most linear when performing the BCH expansion (also labeled with the subscript $L^\prime$).
Exploiting this new notation, eqs.~\eqref{eq:lcc+pccd} and \eqref{eq:lcc+pccd-exp} are therefore equivalent.
\section{Orbital-based correlation measures}\label{sec:i12}
In this work, we will focus on orbital-based entanglement and correlation measure.
Specifically, we will study the single orbital entropy and orbital-pair mutual information.
The single orbital entropy quantifies the entanglement between one particular orbital and the remaining set of orbitals contained in the active orbital space and is defined as \cite{dmrg-10}
\begin{equation}\label{eq:single-entropy_1}
s(1)_i = -\sum_{\alpha=1}^{4} \omega_{\alpha;i}\ln(\omega_{\alpha;i}),
\end{equation}
where $\omega_{\alpha,i}$ are the eigenvalues of the one-orbital reduced density-matrix~(1O-RDM) for orbital $i$. Thus, the single orbital entropy of orbital $i$ is the von Neumann entropy of the corresponding 1O-RDM.
The 1O-RDM can be calculated from the one- and two-particle reduced density matrices (1- and 2-RDM) \cite{boguslawski2015,ijqc-eratum},  $\gamma_q^p$ and $\Gamma_{rs}^{pq}$, respectively.
For variationally optimized wave functions, the 1- and 2-RDM are defined as an expectation value of the form
\begin{equation}\label{eq:1rdm}
\gamma_q^p = \frac{\langle \Psi|a_p^\dagger a_q |\Psi\rangle}{\langle \Psi|\Psi\rangle}
\end{equation}
and
\begin{equation}\label{eq:2rdm}
\Gamma_{rs}^{pq} = \frac{\langle \Psi|a_p^\dagger a_q^\dagger a_s a_r |\Psi\rangle}{\langle \Psi|\Psi\rangle}.
\end{equation}
In terms of the spin-dependent 1- and 2-RDMs, the 1O-RDM $\rho_i$ is diagonal and takes over a rather simple form \cite{boguslawski2015,ijqc-eratum}. For a general wave function, it can be determined as
\begin{equation}\label{eq:lccsdrho1}
\mathbf{\rho}_i =
\left( \begin{array}{cccc}
1-\gamma_i^i-\gamma_{\overline{i}}^{\overline{i}} + \Gamma_{i\overline{i}}^{i\overline{i}} & 0 & 0&0\\
0 & \gamma_i^i - \Gamma_{i\overline{i}}^{i\overline{i}}& 0& 0\\
0 & 0 &\gamma_{\overline{i}}^{\overline{i}} - \Gamma_{i\overline{i}}^{i\overline{i}}&0\\
0& 0 &0 & \Gamma_{i\overline{i}}^{i\overline{i}}
\end{array} \right)
\end{equation}
where the indices $i$ label the orbital in question and $i$ ($\overline{i}$) indicate spin-up (spin-down) electrons.
The above matrix is expressed in the basis \{\zero,\up,\down,\double\},  representing the 4 possible occupations of a (spatial) orbital.
We should note that for pCCD $\rho_i$ can be further simplified.
Since the pCCD model excludes singly-occupied orbitals and we have the relation $\gamma_i^i=\Gamma_{i\overline{i}}^{i\overline{i}}$, the corresponding 1O-RDM of pCCD is a 2x2 matrix represented in the basis \{\zero,\double\} \cite{boguslawski2016},
\begin{equation}\label{eq:pccdrho1}
\mathbf{\rho}_i^{\rm {pCCD}} =
\left(\begin{array}{cc}
1-\gamma_i^i & 0\\
0 & \gamma_i^i
\end{array} \right),
\end{equation}
as all the remaining matrix elements sum up to zero.
When the LCC corrections are applied on top of the pCCD wave function, the overall 1O-RDM will be the sum of its individual contributions,
\begin{equation}\label{eq:pccdlccsdrho1}
\mathbf{\rho}_i^{\rm {pCCD-LCC}}  = \mathbf{\rho}_i^{\rm {pCCD}} +\mathbf{\rho}_i^{\rm {LCC}},
\end{equation}
where $\mathbf{\rho}_i^{\rm {pCCD}}$ has to be expanded to its 4x4 analog.
Therefore, {the hybrid pCCD-LCC 1O-RDM $\mathbf{\rho}_i^{\rm {pCCD-LCC}}$ is} s quadratic matrix of rank 4 spanned by the basis states of the one-orbital Fock space (see also eq.~\eqref{eq:lccsdrho1}).

In an analogous way, the {correlation} between two orbitals $i,j$ and the remaining set of orbitals is measured by the two-orbital entropy,
\begin{equation}\label{eq:single-entropy_2}
s_{i,j} = -\sum_{\alpha=1}^{16} \omega_{\alpha;i,j}\ln(\omega_{\alpha;i,j}),
\end{equation}
where $\omega_{\alpha;i,j}$ are the eigenvalues of the two-orbital (2O-)RDM $\rho_{i,j}$.
In contrast to eq.~\eqref{eq:single-entropy_1}, the 2O-RDM is defined in terms of basis states of a two-orbital Fock space, which contains 16 possible basis states in the case of spatial orbitals.
The matrix elements of the 2O-RDM can be express in terms of the one-, two-, three-, and four-particle RDMs \cite{boguslawski2015,ijqc-eratum}, $\gamma_q^p$, $\Gamma_{rs}^{pq}$, $\Gamma_{stu}^{pqr}$, and $\Gamma_{tuvw}^{pqrs}$, respectively, with
\begin{equation}\label{eq:3rdm}
\Gamma_{stu}^{pqr}= \frac{\langle \Psi|a_p^\dagger a_q^\dagger a_r^\dagger a_u a_t a_s |\Psi\rangle}{\langle \Psi|\Psi\rangle}
\end{equation}
and 
\begin{equation}\label{eq:particle RDM2}
\Gamma_{tuvw}^{pqrs}= \frac{\langle \Psi|a_p^\dagger a_q^\dagger a_r^\dagger  a_s^\dagger a_w a_v a_u a_t |\Psi\rangle}{\langle \Psi|\Psi\rangle}.
\end{equation}
If the electronic wave function is an eigenfunction of $\hat{S}_z$, the 2O-RDM is block-diagonal \cite{entanglement_bonding_2013}. The corresponding matrix elements in terms of the $N$-RDMs are summarized in Table~\ref{tbl:rho2} (see also Refs.~\citenum{entanglement_bonding_2013,boguslawski2015,ijqc-eratum}).
\begin{table*}
\begin{threeparttable}[th]
\centering\small
\caption{$\rho_{i,j}^{(2)}$ expressed in terms of $N$-RDMs. For restricted orbitals, the equivalent sub-blocks are color-coded (see text).}\label{tbl:rho2}
\begin{tabular}{c|cccccccccccccccc}
\hline
    &\zero\,\zero                                                               
    &\zero\,\up                                                                 
    &\up\,\zero                                                                 
    &\zero\,\down                                                               
    &\down\,\zero                                                               
    &\up\,\up                                                                   
    &\down\,\down                                                               
    &\zero\,\double                                                             
    &\up\,\down                                                                 
    &\down\,\up                                                                 
    &\double\,\zero                                                             
    &\up\,\double                                                               
    &\double\,\up                                                               
    &\down\,\double                                                             
    &\double\,\down                                                             
    &\double\,\double   \\[0.05in] \hline
 \zero\,\zero     & (1,1) & 0 & 0 & 0 & 0 & 0 & 0 & 0 & 0 & 0 & 0 & 0 & 0 & 0 & 0 & 0 \\[0.05in]
 \zero\,\up       & 0 & \cellcolor{gray!10} (2,2) & \cellcolor{gray!10} (2,3) & 0 & 0 & 0 & 0 & 0 & 0 & 0 & 0 & 0 & 0 & 0 & 0 & 0 \\[0.05in]
 \up\,\zero       & 0 & \cellcolor{gray!10} (3,2) & \cellcolor{gray!10} (3,3) & 0 & 0 & 0 & 0 & 0 & 0 & 0 & 0 & 0 & 0 & 0 & 0 & 0 \\[0.05in]
 \zero\,\down     & 0 & 0 & 0 & \cellcolor{gray!10}(4,4) & \cellcolor{gray!10}(4,5) & 0 & 0 & 0 & 0 & 0 & 0 & 0 & 0 & 0 & 0 & 0 \\[0.05in]
 \down\,\zero     & 0 & 0 & 0 & \cellcolor{gray!10}(5,4) & \cellcolor{gray!10}(5,5) & 0 & 0 & 0 & 0 & 0 & 0 & 0 & 0 & 0 & 0 & 0 \\[0.05in]
 \up\,\up         & 0 & 0 & 0 & 0 & 0 & \cellcolor{gray!30}(6,6) & 0 & 0 & 0 & 0 & 0 & 0 & 0 & 0 & 0 & 0 \\[0.05in]
 \down\,\down     & 0 & 0 & 0 & 0 & 0 & 0 & \cellcolor{gray!30}(7,7) & 0 & 0 & 0 & 0 & 0 & 0 & 0 & 0 & 0 \\[0.05in]
 \zero\,\double   & 0 & 0 & 0 & 0 & 0 & 0 & 0 & (8,8) & (8,9) & (8,10) & (8,11) & 0 & 0 & 0 & 0 & 0 \\[0.05in]
 \up\,\down       & 0 & 0 & 0 & 0 & 0 & 0 & 0 & (9,8) & (9,9) & (9,10) & (9,11) & 0 & 0 & 0 & 0 & 0 \\[0.05in]
 \down\,\up       & 0 & 0 & 0 & 0 & 0 & 0 & 0 & (10,8) & (10,9) & (10,10) & (10,11) & 0 & 0 & 0 & 0 & 0 \\[0.05in]
 \double\,\zero   & 0 & 0 & 0 & 0 & 0 & 0 & 0 & (11,8) & (11,9) & (11,10) & (11,11) & 0 & 0 & 0 & 0 & 0 \\[0.05in]
 \up\,\double     & 0 & 0 & 0 & 0 & 0 & 0 & 0 & 0 & 0 & 0 & 0 & \cellcolor{gray!75}(12,12) & \cellcolor{gray!75}(12,13) & 0 & 0 & 0 \\[0.05in]
 \double\,\up     & 0 & 0 & 0 & 0 & 0 & 0 & 0 & 0 & 0 & 0 & 0 & \cellcolor{gray!75}(13,12) & \cellcolor{gray!75}(13,13) & 0 & 0 & 0 \\[0.05in]
 \down\,\double   & 0 & 0 & 0 & 0 & 0 & 0 & 0 & 0 & 0 & 0 & 0 & 0 & 0 & \cellcolor{gray!75}(14,14) & \cellcolor{gray!75}(14,15) & 0 \\[0.05in]
 \double\,\down   & 0 & 0 & 0 & 0 & 0 & 0 & 0 & 0 & 0 & 0 & 0 & 0 & 0 & \cellcolor{gray!75}(15,14) & \cellcolor{gray!75}(15,15) & 0 \\[0.05in]
 \double\,\double & 0 & 0 & 0 & 0 & 0 & 0 & 0 & 0 & 0 & 0 & 0 & 0 & 0 & 0 & 0 & (16,16) \\[0.05in]
\hline
\hline
\end{tabular}
\begin{tablenotes}\footnotesize
\item $(2,3) = (3,2)^\dagger = \gamma_{i}^{j}-\Gamma_{i \ib}^{j \ib}-\Gamma_{i \jb}^{j \jb}+{}^3\Gamma_{i \ib \jb}^{j \ib \jb}$
\item $(4,5)=(5,4)^\dagger = \gamma_{\ib}^{\jb}-\Gamma_{i \ib}^{i \jb}-\Gamma_{j \ib}^{j \jb} + {}^3\Gamma_{ij\ib}^{ij\jb}$
\item $(6,6)= \Gamma_{ij}^{ij}-{}^3\Gamma_{i\ib j}^{i\ib j}-{}^3\Gamma_{ij\jb}^{ij\jb}+{}^4\Gamma_{i\ib j \jb}^{i\ib j \jb}$
\item $(7,7)= \Gamma_{\ib \jb}^{\ib \jb}-{}^3\Gamma_{i \ib \jb}^{i\ib \jb}-{}^3\Gamma_{\ib j\jb}^{\ib j\jb}+{}^4\Gamma_{i\ib j \jb}^{i\ib j \jb}$
\item $(8,8)= \Gamma_{j\jb}^{j\jb}-{}^3\Gamma_{i j\jb}^{ij\jb}-{}^3\Gamma_{\ib j\jb}^{\ib j\jb}+{}^4\Gamma_{i\ib j \jb}^{i\ib j \jb}$
\item $(8,9)=(9,8)^\dagger = \Gamma_{i\jb}^{j\jb}-{}^3\Gamma_{i\jb \ib}^{j \jb \ib}$
\item $(8,10)=(10,8)^\dagger = -\Gamma_{j\ib}^{j\jb}+{}^3\Gamma_{ij \ib}^{i j \jb}$
\item $(8,11)=(11,8)^\dagger = \Gamma_{i\ib}^{j\jb}$
\item $(9,9)= \Gamma_{i\jb}^{i\jb}-{}^3\Gamma_{i\ib \jb}^{i\ib \jb}-{}^3\Gamma_{i j \jb}^{i j \jb}+{}^4\Gamma_{i\ib j\jb}^{i\ib j\jb}$
\item $(9,10)=(10,9)^\dagger = -\Gamma_{j\ib}^{i\jb}$
\item $(9,11)=(11,9)^\dagger = \Gamma_{i\ib}^{i\jb}-{}^3\Gamma_{ij\ib}^{ij\jb}$
\item $(10,10)= \Gamma_{\ib j}^{\ib j}-{}^3\Gamma_{i\ib j}^{i\ib j}-{}^3\Gamma_{\ib j \jb}^{\ib j \jb}+{}^4\Gamma_{i\ib j\jb}^{i\ib j\jb}$
\item $(10,11)=(11,10)^\dagger = -\Gamma_{i\ib}^{j\ib}+{}^3\Gamma_{i\ib \jb}^{j\ib \jb}$
\item $(11,11)= \Gamma_{i\ib}^{i\ib}-{}^3\Gamma_{i\ib j}^{i \ib j}-{}^3\Gamma_{i \ib \jb}^{i \ib \jb}+{}^4\Gamma_{i\ib j\jb}^{i\ib j\jb}$
\item $(12,12)= {}^3\Gamma_{ij\jb}^{ij\jb}-{}^4\Gamma_{i\ib j\jb}^{i\ib j\jb}$
\item $(12,13)= (13,12)^\dagger = -{}^3\Gamma_{ij\ib}^{ij\jb}$
\item $(13,13)= {}^3\Gamma_{i\ib j}^{i\ib j}-{}^4\Gamma_{i\ib j\jb}^{i\ib j\jb}$
\item $(14,14)= {}^3\Gamma_{\ib j \jb}^{\ib j \jb}-{}^4\Gamma_{i\ib j\jb}^{i\ib j\jb}$
\item $(14,15)= (15,14)^\dagger = -{}^3\Gamma_{\ib i \jb}^{\ib j \jb}$
\item $(15,15)= {}^3\Gamma_{i\ib \jb}^{i\ib \jb}-{}^4\Gamma_{i\ib j\jb}^{i\ib j\jb}$
\item $(16,16)= {}^4\Gamma_{i\ib j\jb}^{i\ib j\jb}$
\item $(1,1) = 1-\gamma_{i}^{i}-\gamma_{\ib}^{\ib}-\gamma_{j}^j-\gamma_{\jb}^{\jb}+\Gamma_{i\ib}^{i\ib}+\Gamma_{j\jb}^{j\jb}+\Gamma_{ij}^{ij}+\Gamma_{i\jb}^{i\jb}+\Gamma_{\ib j}^{\ib j}+\Gamma_{\ib\jb}^{\ib\jb}-{}^3\Gamma_{ij\jb}^{ij\jb}-{}^3\Gamma_{\ib j\jb}^{\ib j\jb}-{}^3\Gamma_{i\ib j}^{i \ib j}-{}^3\Gamma_{i\ib\jb}^{i\ib\jb}+{}^4\Gamma_{i\ib j\jb}^{i \ib j\jb}$
\item $(2,2) = \gamma_{j}^{j}-\Gamma_{ij}^{ij}-\Gamma_{\ib j}^{\ib j}-\Gamma_{j \jb}^{j \jb}+{}^3\Gamma_{i\jb j}^{i\jb j}+{}^3\Gamma_{i \ib j}^{i \ib j}+{}^3\Gamma_{\ib j\jb}^{\ib j\jb}-{}^4\Gamma_{i \ib j \jb}^{i \ib j \jb}$\\
\item $(3,3)= \gamma_i^i-\Gamma_{i \ib}^{i \ib} -\Gamma_{ij}^{ij} - \Gamma_{i\jb}^{i\jb}+{}^3\Gamma_{ij\jb}^{ij\jb}+{}^3\Gamma_{i\ib j}^{i\ib j} +{}^3\Gamma_{i\ib \jb}^{i\ib \jb} -{}^4\Gamma_{i\ib j\jb}^{i \ib j \jb}$
\item $(4,4)=\gamma_{\jb}^{\jb}-\Gamma_{i\jb}^{i \jb}-\Gamma_{\ib \jb}^{\ib \jb}-\Gamma_{j\jb}^{j\jb}+{}^3\Gamma_{i \ib \jb}^{i \ib \jb}+{}^3\Gamma_{ij\jb}^{ij\jb}+{}^3\Gamma_{\ib j \jb}^{\ib j \jb}-{}^4\Gamma_{i\ib j\jb}^{i\ib j \jb}$
\item $(5,5)=\gamma_{\ib}^{\ib}-\Gamma_{\ib j}^{\ib j}-\Gamma_{\ib \jb}^{\ib \jb}-\Gamma_{i\ib}^{i\ib}+{}^3\Gamma_{\ib j \jb}^{\ib j \jb}+{}^3\Gamma_{i\ib j}^{i \ib j}+{}^3\Gamma_{i \ib \jb}^{i \ib \jb}-{}^4\Gamma_{i\ib j\jb}^{i\ib j \jb}$
\end{tablenotes}
\end{threeparttable}
\end{table*}
As for $\rho_{i}^{\rm pCCD}$, all singly-occupied states have vanishing matrix elements and the 2O-RDM of pCCD reduces to a 4x4 matrix represented in the basis states  \{\zero\,\zero,\double\,\zero,\zero\,\double,\double\,\double\} \cite{boguslawski2016},
\begin{equation}\label{eq:pccdrho2}
\mathbf{\rho}_{i,j}^{\rm{pCCD}} =
\left( \begin{array}{cccc}
1-\gamma_i^i-\gamma_{{j}}^{{j}} + \Gamma_{i\overline{j}}^{i\overline{j}} & 0 & 0&0\\
0 & \gamma_j^j - \Gamma_{i\overline{j}}^{i\overline{j}}& \Gamma^{j\overline{j}}_{i\overline{i}}& 0\\
0 & \Gamma^{i\overline{i}}_{j\overline{j}} &\gamma_{i}^{i} - \Gamma_{i\overline{j}}^{i\overline{j}}&0\\
0& 0 &0 & \Gamma_{i\overline{j}}^{i\overline{j}}
\end{array} \right)
\end{equation}
The 2O-RDM $\rho_{i,j}^{\rm pCCD-LCC}$ of pCCD-LCC is again the sum of its individual contributions, summarized in Table~\ref{tbl:rho2} and eq.~\eqref{eq:pccdrho2},
\begin{equation}\label{eq:rho2pccd-lccsd}
\rho_{i,j}^{\rm pCCD-LCC} = \rho_{i,j}^{\rm pCCD} + \rho_{i,j}^{\rm LCC}.
\end{equation}
From $s_i$ and $s_{i,j}$, we can determine the orbital-pair mutual information \cite{rissler2006}
\begin{equation}\label{eq:mutal info}
I_{i|j} = s_i + s_j - s_{i,j},
\end{equation}
which measures the correlation between the orbital pair $i$ and $j$ embedded in the environment of all other active-space orbitals.

The calculation of the 1O- and 2O-RDM for pCCD has been investigated in Ref.~\cite{boguslawski2016}.
In the following, we will thus focus on how the contributions of each LCC correction to the 1O- and 2O-RDM of the hybrid pCCD-LCC method can be determined.
\section{Representation of correlation measures in terms of LCC response density matrices}\label{sec:i12-lcc}
One possibility to evaluate the 1O- and 2O-RDM of {pCCD-LCC} is to determine the response density matrices of pCCD \cite{boguslawski2016} and the {LCC corrections.}
Specifically, the correlation contribution of the CC correction on top of the pCCD wave function can be determined from
\begin{align}\label{eq:responsedms-corr}
(\Gamma^{p\ldots}_{t\ldots})^{\rm corr}=\bra{\Phi_0}(1+\Lambda^\prime) & \{ e^{-\hat{T}^\prime -\hat{T}_p} \{\hat{a}^\dagger_p\ldots a_t \}  \nonumber \\
&e^{\hat{T}_p+\hat{T}^\prime} \}_{L^\prime} \ket{\Phi_0},
\end{align}
where $\Lambda^\prime=\Lambda_1+\Lambda_2^\prime$ or $\Lambda^\prime=\Lambda_2^\prime$, respectively, and
\begin{equation}\label{eq:de-excitation}
\Lambda_n^\prime = {{1}\over{(n!)^2}}\sum_{ij...}\sum_{ab...}{}^\prime \lambda^{ij...}_{ab...}{i^\dagger a j^\dagger b...}
\end{equation}
is the de-excitation operator, where all electron-pair de-excitation are to be excluded as they do not enter the {LCC} equations (again, indicated by the "$\prime$").
Note that we have written eq.~\eqref{eq:responsedms-corr} using the exponential ansatz introduced in eq.~\eqref{eq:lcc+pccd-exp}.
When evaluating the {LCC} response density matrices, only terms that are at most linear in $\hat{T}_1$ and $\hat{T}_2^\prime$ are to be considered.
The total $N$-RDM used to construct the 1O- and 2O-RDMs is the sum of the reference contribution, the leading correlation contribution eq.~\eqref{eq:responsedms-corr}, and all lower-order correlation contributions,  
\begin{equation}\label{eq:responsedms}
\Gamma^{p\ldots}_{t\ldots}= (\Gamma^{p\ldots}_{t\ldots})^{\rm ref}+(\Gamma^{p\ldots}_{t\ldots})^{\rm corr} + \{(\Gamma^{p\ldots}_{t\ldots})^{\rm corr}_{(N-1,\ldots,1)}\} ,
\end{equation}
where the last term indicates all possible lower-order ($N-1,\ldots,1$) correlation contributions to the $N$-RDM in question.

\begin{figure}[tp]
\includegraphics[width=0.5\textwidth]{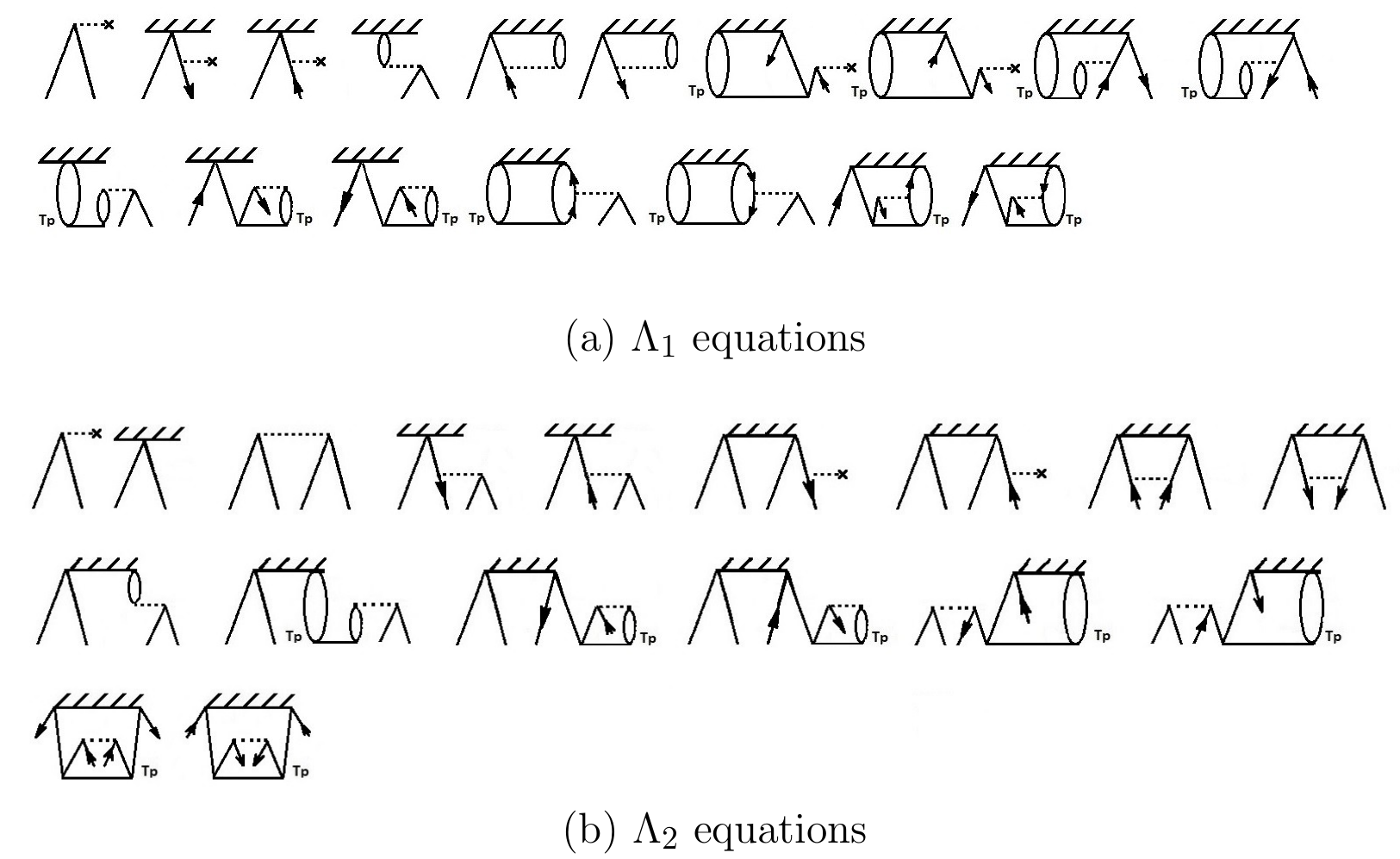}
\caption{Diagrammatic representation of the pCCD-LCCSD $\Lambda$ equations using antisymmetrized Goldstone diagrams (second line of eq.~\eqref{eq:lambda}). $\hat T_{\rm p}$ indicates electron-pair excitation vertices. For pCCD-LCCD, the $\Lambda_2$ equations do not contain any terms comprising $\Lambda_1$ vertices.}\label{fig:lambda-eq}
\end{figure}

To evaluate the response density matrices, the $\{\lambda^{ij...}_{ab...}\}$ amplitudes are required, which are determined by solving the $\Lambda$ equations of the LCC corrections, where the first derivative of the LCC Lagrangian
\begin{align}\label{eq:energy}
\mathcal{L}=E(\lambda, t) = \langle \Phi_0|(1+\Lambda^\prime) \{ &e^{-\hat{T}^\prime -\hat{T}_p}\hat{H} \nonumber \\
&e^{\hat{T}_p+ \hat{T}^\prime} \}_{L^\prime}|\Phi_0\rangle,
\end{align}
with respect to the LCC amplitudes $t_\nu$ (again, excluding all electron pairs) have to vanish
\begin{align}\label{eq:lambda}
\frac{\partial \mathcal{L}}{\partial t_\nu}   &= \frac{\partial }{\partial t_\nu} \langle \Phi_0|(1+\Lambda^\prime)  \{e^{-\hat{T}^\prime -\hat{T}_p}\hat{H}e^{\hat{T}_p +\hat{T}^\prime}\}_{L^\prime}|\Phi_0\rangle \nonumber \\
&= \langle \Phi_0|(1+\Lambda^\prime) \{e^{-\hat{T}^\prime -\hat{T}_p} [\hat{H},\hat{\tau}_\nu] e^{\hat{T}_p +\hat{T}^\prime} \}_{L^\prime}^\prime|\Phi_0\rangle \nonumber \\
&= 0.
\end{align}
Due to the linear ansatz of the CC correction, the resulting $\Lambda$ equations have a particular simple form and can be solved efficiently.
Specifically, the only excitation vertices that couple to $\hat{\tau}_\nu$ include $\hat{T}_{\rm p}$.
This simplification due to the partial derivative with respect to the broken-pair amplitudes has been indicated by the superscript ``$\prime$'' around the curly brackets.
The diagrammatic representation of the LCCD and LCCSD $\Lambda$ equations of pCCD-LCCD and pCCD-LCCSD are shown in Fig.~\ref{fig:lambda-eq}, while their algebraic representation is summarized in eq.~(1) of the Supporting Information.
We should note that we do not consider any orbital response contributions in the LCC $\Lambda$ equations.
Thus, the resulting $N$-RDMs are unrelaxed as they do not account for orbital relaxation effects.
The pCCD RDMs, on the other hand, are relaxed as they are determined from an orbital-optimized pCCD reference function.

Once the $\Lambda$ equations of the {LCC corrections} have been solved, all required elements of the 1-, 2-, 3-, and 4-RDM can be evaluated (using eqs.~\eqref{eq:responsedms-corr} and \eqref{eq:responsedms}). Note that the 1O- and 2O-RDMs are expressed using spin-resolved $N$-RDMs and thus spin-free RMDs cannot be applied to construct the entanglement and correlation measures.
This generalization does, however, not pose any difficulties even if spin-free CC calculations have been performed, where all CC amplitudes are spin-summed.
When evaluating eq.~\eqref{eq:responsedms} or eq.~\eqref{eq:responsedms-corr}, we can only consider the spin-block in question (e.g., $a^\dagger_{\bar{p}}a_{\bar{p}}$, etc.) omitting the spin-summation step.

Furthermore, the resulting 1O- and 2O-RDM can be simplified if restricted orbitals are used, where spin-up and spin-down electrons have a common set of spatial orbitals.
The resulting 1O-RDM contains only 3 distinct elements (one for \zero, \double, or \up/\down), while the 2O-RDM comprises only 6 distinct sub-blocks.
The equivalent sub-blocks are marked in gray in Table~\ref{tbl:rho2}, while all unique elements of the pCCD-LCC 2O-RDM are summarized in Table S2 of the Supporting Information.
Finally, the procedure to determine the orbital entanglement and correlation measure for any pCCD-CC model is similar as scrutinized above, where the similarity transformed Hamiltonian (including the de-excitation operator, if required) has to be adjusted according to the chosen flavor of the CC correction.

\section{Computational details}\label{sec:details}
For the {investigated molecules}, we have employed Dunning’s cc-pVDZ basis set with the following contraction (9s,4p,1d) $\rightarrow$ [3s,2p,1d] \cite{basis_dunning}.

In all pCCD calculations, we have applied the variational orbital optimization protocol presented
in Ref.~\cite{oo-ap1rog} to minimize the size-consistency errors due to the stretching of molecular bonds.
All pCCD-based calculations, including the orbital optimization, have been performed using the \textsc{PyBEST} software package  \cite{piernik100}.

As reference, we have performed density-matrix renormalization group (DMRG) calculations with the BUDAPEST QC-DMRG program \cite{dmrg_ors}.
The molecular orbitals obtained from the orbital-optimized pCCD calculations have been chosen as the orbital basis in all DMRG calculations. The orbital ordering was optimized to enhance DMRG convergence. The minimum and maximum number of renormalized active-system states $m$ was varied from 256 to 1024, respectively. Finally, we have also employed the dynamic block state selection (DBSS) protocol \cite{legeza_ors_04} to dynamically choose the number of renormalized active-system states according to a predefined threshold value for the quantum information loss. Specifically, the minimum value for $m$ was set to 1024, the maximum value to 2048, while the threshold for the quantum information loss was set to $10^{-5}$ in all DBSS-based DMRG calculations.   



\section{Orbital-pair correlation spectra}\label{sec:results}
As a proof of principal analysis, we have chosen three small, but challenging test systems where the performance of pCCD is modest.
These include systems dominated by static/nondynmaic correlation (\ce{N2}), dynamical correlation (\ce{F2}), and the one-dimensional half-filled Hubbard Hamiltonian, where pCCD provides a poor reference function in the strong correlation limit.

\begin{figure*}[tp]
  \includegraphics[width=1.0\textwidth]{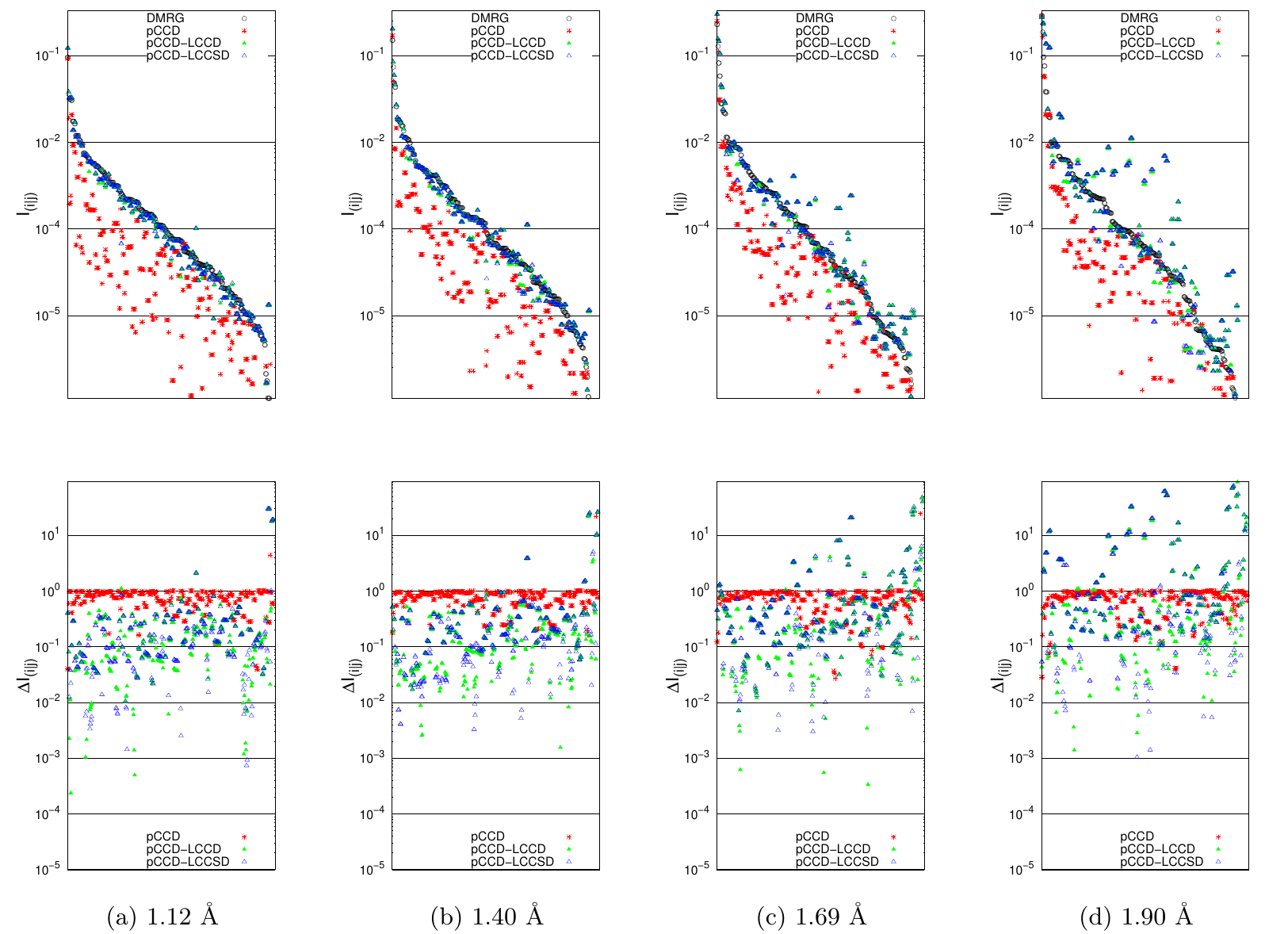}
\caption{The decay of the orbital-pair mutual information (upper panel) and its relative absolute difference with respect to the DMRG reference distribution (lower panel) for the \ce{N2} molecule at different points along the reaction coordinate. The decay of orbital-pair mutual information is sorted with respect to the DMRG reference distribution. Thus each point $I_{i|j}$ corresponds to the same orbital pair $i,j$.
The relative abosule difference is calculated as $\Delta I_{i|j} = \frac{|I_{i|j}(\textrm{method})-I_{i|j}(\textrm{DMRG})|}{I_{i|j}(\textrm{DMRG})}$.}\label{fig:n2-decay}
\end{figure*}

\subsection{The nitrogen dimer}\label{sec:n2}
The accurate quantum chemical description of the dissociation process of the \ce{N2} molecule is a challenging problem for state-of-the-art computational chemistry.
Specifically, along the dissociation pathway, the orbital correlation spectrum changes: while the nitrogen dimer is dominated by dynamical correlation around the equilibrium geometry, the contribution of nondynamic/static correlation significantly increases when the N--N bond is stretched.
Most importantly, conventional CC methods, including pCCD-based approaches, break down in the dissociation limit \cite{chan_n2,entanglement_letter,boguslawski2017} (for $r_{\rm{NN}} > 2.00$ \AA{}, depending on the chosen (truncated) CC model).
This breaking point has thus to be visible in the correlation measures.
To study the orbital-pair correlation spectra along the dissociation pathway, we have chosen four different points along the reaction coordinate raging from the equilibrium distance to the breaking point of pCCD-LCC. The total electronic energies determined for various quantum chemistry methods are summarized in Table~S1.1 in the Supporting Information. 

Fig.~\ref{fig:n2-decay} displays the decay of the orbital-pair correlation spectra obtained for pCCD, pCCD-LCCD, pCCD-LCCSD, and DMRG at selected interatomic distances (upper panel) and the relative absolute differences in the orbital-pair mutual information with respect to the DMRG reference spectrum (lower panel).
In general, the LCC correction on top of pCCD significantly improves the orbital-pair correlation spectrum at each of the selected interatomic distances (see also Fig.~S10 in the Supporting Information).
Specifically, both pCCD-LCCD and pCCD-LCCSD accurately recover the moderate/weak orbital-pair correlations ($I_{ij}<10^{-1}$) around equilibrium and for a stretched N--N bond.
The LCCD and LCCSD corrections, however, fail for strongly-correlated orbital pairs
($I_{ij}>10^{-1}$), where the corresponding values of the mutual information are overestimated compared to the DMRG reference data.
If we approach the breaking point of pCCD-LCC ($r_{\rm{NN}}\approx 2.00$ \AA{}), the LCC corrections overestimate all orbital-pair correlations compared to the DMRG reference spectrum (see Fig.~\ref{fig:n2-decay}(d)).
This overcorrelation is also evident in the total electronic energies (see Table S1 of the Supporting Information).
While the differences in total energies are small around the equilibrium (around $1\,mE_h$), they significantly increase for a stretched N--N bond (up to $10\,mE_h$).
The overestimation of the correlation energy can be attributed to the linearized ansatz of the CC correction. 

To conclude, the LCC corrections on top of pCCD significantly improves the description of weak electron correlation effects, where the predicted values of the orbital-pair mutual information agree well with the DMRG reference (for both the equilibrium geometry and stretched bonds).
Nonetheless, pCCD-LCC overestimates the dominant orbital-pair correlations along the whole dissociation pathway, which results in too low total energies.



\begin{figure*}[tbp]
  \includegraphics[width=1.0\textwidth]{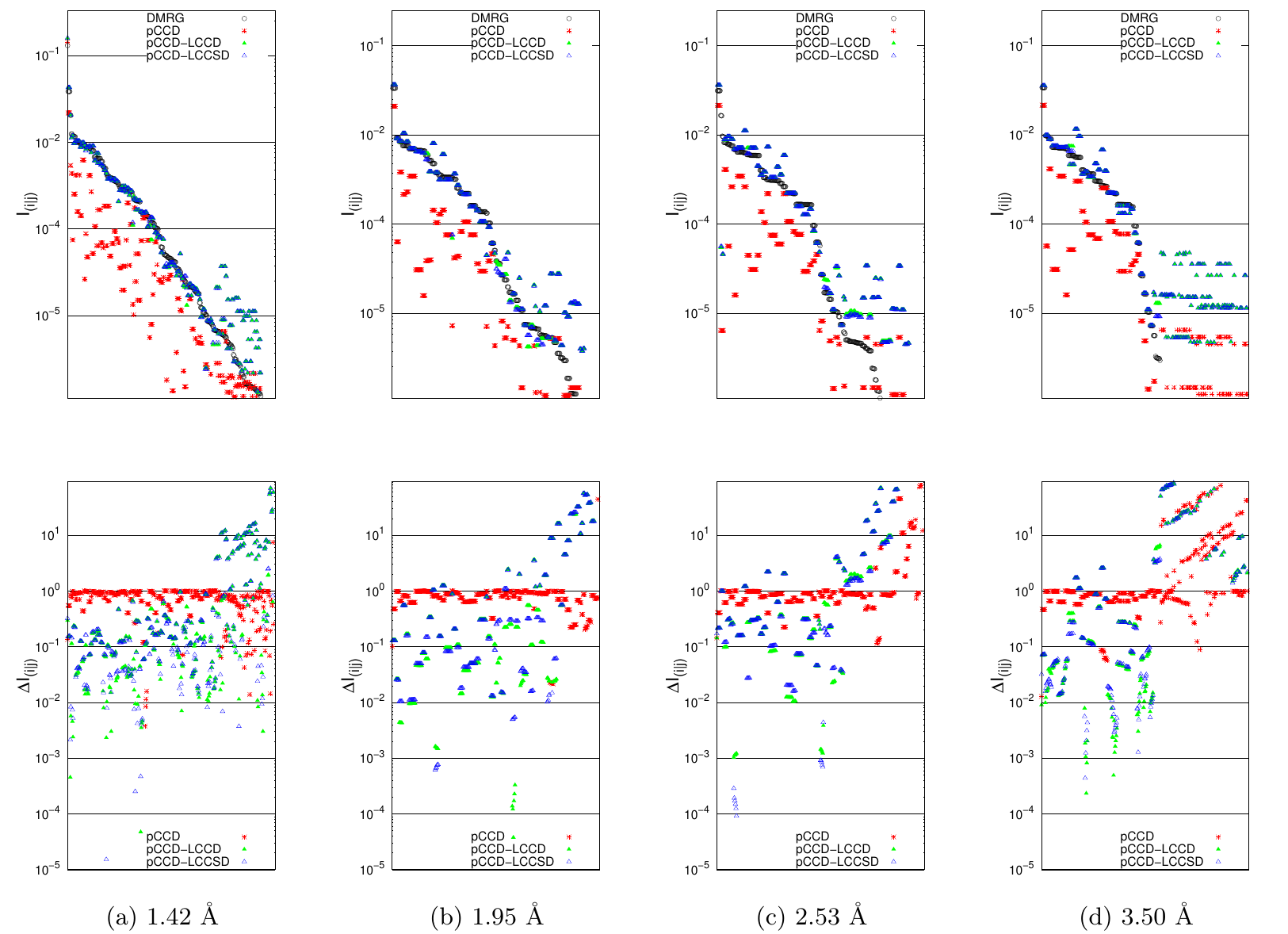}
\caption{
The decay of the orbital-pair mutual information (upper panel) and its relative absolute difference with respect to the DMRG reference distribution (lower panel) for the \ce{F2} molecule at different points along the reaction coordinate. The decay of orbital-pair mutual information is sorted with respect to the DMRG reference distribution. Thus each point $I_{i|j}$ corresponds to the same orbital pair $i,j$.
The relative abosule difference is calculated as $\Delta I_{i|j} = \frac{|I_{i|j}(\textrm{method})-I_{i|j}(\textrm{DMRG})|}{I_{i|j}(\textrm{DMRG})}$.}\label{fig:f2decay}
\end{figure*}

\subsection{The fluorine dimer}\label{sec:f2}
In contrast to \ce{N2}, the \ce{F_2} molecule is dominated by dynamical correlation.
Furthermore, in order to describe the dissociation process of the fluorine dimer correctly, triple (or higher-order) excitations are important \cite{kowalski-f2_2001}.
Conventional CC models restricted to double excitations, thus, predict spectroscopic constants of rather poor accuracy compared to more sophisticated methods and/or experimental results \cite{kowalski-f2_2001,boguslawski2017}.
The dissociation process of the \ce{F2} molecule, hence, represents a good test system to scrutinize orbital-pair correlations in pCCD-based methods, which are restricted to at most double excitations.
As in the case of the nitrogen molecule, we will focus our analysis on selected points along the potential energy surface.
The corresponding energies are summarized in the Supporting Information.

Fig.~\ref{fig:f2decay} displays the decay of the orbital-pair correlation spectra obtained for pCCD, pCCD-LCCD, pCCD-LCCSD, and DMRG at selected interatomic distances (upper panel) and the relative absolute differences in the orbital-pair mutual information with respect to the DMRG reference spectrum (lower panel).
As observed for the \ce{N2} molecule, the LCCD and LCCSD corrections on top of pCCD improve the orbital-pair correlation spectrum for each point of the reaction coordinate, where both LCC flavours recover the missing dynamical correlation effects in the pCCD reference function (see also section S7 of the Supporting Information, which summarizes the eigenvalue spectra of the 2O-RDM).
Most importantly, the linearized CC corrections are able to capture the essential part of the dynamical correlation around equilibrium, while the leading orbital-pair correlations are slightly overestimated.
However, the performance of pCCD-LCC slightly deteriorates for stretched molecular bonds.
From 1.95 \AA{} onwards, the weakest orbital-pair correlations ($I_{i|j} < 10^{-4}$) are overestimated compared to the DMRG reference results.
Around the equilibrium geometry as well as for stretched distances, both pCCD-LCC methods predict lower total energies than DMRG (differences lie between 2.00 and 9.00 $mE_h$, see also Tabel S1.2 of the Supporting Information).

To conclude, both pCCD-LCCD and pCCD-LCCSD accurately recover the orbital-pair correlation spectrum of the \ce{F2} molecule around equilibrium and in the vicinity of dissociation.
Their performance, however, deteriorates when the F--F bond is moderately stretched.
Nonetheless, the LCC corrections are able to cure some of the deficiencies of the pCCD model resulting in overall smaller relative absolute differences along the reaction coordinate.



\begin{figure*}[tbp]
  \includegraphics[width=1.0\textwidth]{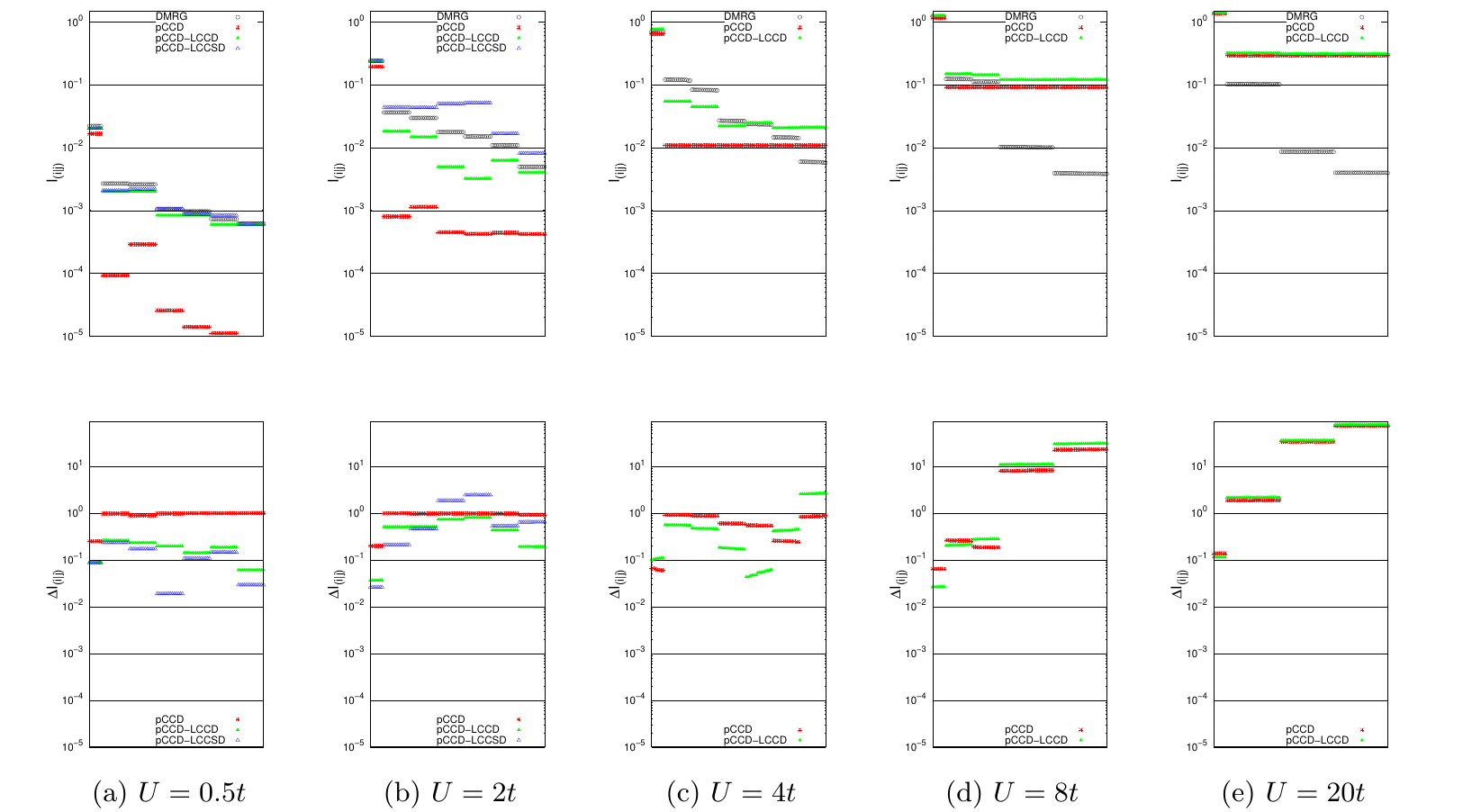}
\caption{
The decay of the orbital-pair mutual information (upper panel) and its relative absolute difference with respect to the DMRG reference distribution (lower panel) for the half-filled Hubbard model with periodic boundary conditions for different on-site interaction strengths. The decay of orbital-pair mutual information is sorted with respect to the DMRG reference distribution. Thus each point $I_{i|j}$ corresponds to the same orbital pair $i,j$.
The relative abosule difference is calculated as $\Delta I_{i|j} = \frac{|I_{i|j}(\textrm{method})-I_{i|j}(\textrm{DMRG})|}{I_{i|j}(\textrm{DMRG})}$.
Note that pCCD-LCCSD does not yield a physical orbital-pair correlation spectrum for $U/t \ge 4$.}\label{fig:hubbard-decay}
\end{figure*}
\subsection{The half-filled one-dimensional Hubbard Hamiltonian}\label{sec:hubbard}
Finally, we will scrutinize the orbital-pair correlation spectra of the one-dimensional (1D) Hubbard Hamiltonian \cite{hubbard_model} with periodic boundary conditions,
\begin{equation}\label{eq:hubbard}
\hat{H}_{\rm Hub} =-t\sum_{\substack{j\\ \sigma \in\{\uparrow,\downarrow\}}}(a^{\dagger}_{j\sigma}a_{(j+1)\sigma} +a^{\dagger}_{(j+1)\sigma}a_{j\sigma}) + U\sum_{j}n_{j\uparrow}n_{j\downarrow},
\end{equation}
where the first part describes the nearest-neighbor hopping, while the second term represents the repulsive on-site interaction, that is, $U>0$.
For the case with $U=0t$ and the local site basis, the half-filled 1D Hubbard model is gapless, where all four local basis states $(\ket{\empty},\ket{\uparrow},\ket{\downarrow},\ket{\uparrow\downarrow})$ have equal weights of $\frac{1}{4}$ and the single-site entropy is $s_i=\ln(4)$.
For increasing $U/t$, the charge gap opens and the weights of the unoccupied and doubly occupied basis states gradually decrease.
In the limit $U/t \rightarrow\infty$, only the singly-occupied states $(\ket{\uparrow},\ket{\downarrow})$ have weights of $\frac{1}{2}$ resulting in a single-site entropy of $s_i=\ln(2)$ as the model becomes equivalent to the spin-$\nicefrac{1}{2}$ Heisenberg model with an antiferromagnetic ground state.
Therefore, wave functions restricted to the seniority-zero sector are not suitable to accurately describe quantum states for an increasing ratio $U/t$, assuming we work in the local on-site basis.
The contributions of the singly-occupied states can be, however, reduced if the site basis is optimized (here, within pCCD).
The corresponding basis optimization results in a localized dimer basis \cite{oo-ap1rog}, where contributions of the unoccupied and doubly-occupied states to the 1O- and 2O-RDM dominate \cite{boguslawski2016}.
However, this localized dimer basis, which is energetically optimal for the pCCD model, yields large single-site entropies of $s_i=\ln(4)$ in the strong correlation limit.
Furthermore, the single-site entropy (see also Fig.~S3 of the Supporting Information) exceeds the strong correlation limit of $s_i=\ln(2)$ of the on-site basis already for $U/t \ge 4$, which highlights the strong multireference nature of the 1D Hubbard model in the pCCD dimer basis.
Compared to the local on-site basis, the pCCD dimer basis will be thus suboptimal in DMRG and, most likely, in post-pCCD calculations.
The 1D Hubbard model, therefore, represents an ideal test case to scrutinize the performance and breaking point of post-pCCD methods.

Similar to Ref.~\citenum{boguslawski2016}, we will focus our discussion on the 1D Hubbard model with 14 sites as similar conclusions can be drawn for longer chain lengths.
The total energies obtained by DMRG, pCCD, pCCD-LCCD, and pCCD-LCCSD are summarized in Table~S3 of the Supporting Information.
As presented in Ref.~\citenum{boguslawski2016}, pCCD underestimates orbital-pair correlations for the small $U/t$ and overestimates them starting from $U/t>2$.
In the strong-correlation limit, pCCD does not provide a reliable orbital-pair correlation spectrum, where all values of the orbital-pair mutual information are overrated.
Fig.~\ref{fig:hubbard-decay} displays the decay of the orbital-pair correlation spectra obtained for pCCD, pCCD-LCCD, pCCD-LCCSD, and DMRG for different on-site interaction strengths (upper panel) and the relative absolute differences in the orbital-pair mutual information with respect to the DMRG reference spectrum (lower panel).
In the weak correlation limit ($U/t<2$), the LCC correction qualitatively improves the orbital-pair correlation spectra, which are similar to the DMRG reference spectrum (see also Fig.{~S12 in the Supporting Information}). Note that LCCD underestimates orbital-pair correlations, while LCCSD overestimates them compared to the DMRG reference distribution.
Most importantly, the LCCSD correction breaks down for $U/t\geq4$, where some of the eigenvalues of the 2O-RDM become negative.
The reason for the break-down of pCCD-LCCSD can be attributed to the overcorrelation in the pCCD reference function and in conjunction with it the large correction predicted by the LCCSD model: when combining the pCCD and LCCSD blocks in the 2O-RDM, the corresponding $N$-RDMs contain large negative values that overshoot the (uncorrelated, but positive) reference part.
Note that the main correction to the energy originates from single excitations (or from the coupling to the single excitation manifold) and therefore pCCD-LCCD does not break in the strong correlation limit.
Thus, the corresponding LCCSD wave functions and total energies for $U/t\geq4$ cannot be considered to be reliable, while the corresponding orbital-pair correlation spectra cannot be unambiguously determined.
We should emphasize that after the breaking point of pCCD-LCCSD ($U/t=4$), the total electronic energies are above the exact energy and the failure of pCCD-LCCSD is only visible in the unphysical orbital-pair correlation spectra.
The performance of pCCD-LCCD is more robust.
Specifically, the LCCD correction provides a reliable descriptions for the breaking point of $U/t=4$.
As expected, for larger values of $U/t$, the overcorrelation introduced by the pCCD reference function cannot be cured by the LCCD correction.
Thus, the pCCD-LCCD correlation spectra are similar to the pCCD ones.

To conclude, although pCCD-LCC fails already for moderate values of $U/t$, it predicts a qualitatively correct step-decay of the orbital-pair correlation spectrum (see Fig.~\ref{fig:hubbard-decay}).
Most importantly, a linearized CC correction on top of pCCD does provide quantitatively reliable orbital-pair correlation spectra in the weak correlation limit.
 

\section{Conclusions}\label{sec:conclusions}
In this work, we investigated the quality of different linearized CC corrections on top of an pCCD reference function using concepts of quantum information theory.
Specifically, we derived the expressions for the 1O- and 2O-RDMs of pCCD-LCCD and pCCD-LCCSD in terms of theirs Lagrangian $N$-particle RDMs and scrutinized the orbital-pair correlation spectra predicted by pCCD, pCCD-LCCD, and pCCD-LCCSD.
Furthermore, we compared the orbital-pair mutual information determined from pCCD-LCC to DMRG reference values (for the same molecular orbital basis).
For our proof-of-principal analysis, we chose the \ce{N_2} (static/nondynamic correlation) and the \ce{F_2} (dynamical correlation) di-atomic molecules and the 1D Hubbard model with periodic boundary conditions and 14 sites.

In general, the LCC corrections predict accurate and reliable orbital-pair correlation spectra around the equilibrium geometry (for molecules) or for the weak correlation limit (for the 1D Hubbard model), where the orbital-pair mutual information lies almost on top of the DMRG reference distribution.
For stretched interatomic distances, pCCD-LCC methods, however, overestimates (nondynamic \textit{and} dynamic) orbital-pair correlations.
Nonetheless, the orbital-pair correlation profile predicted by both pCCD-LCCD and pCCD-LCCSD agrees with the DMRG reference distribution qualitatively, while pCCD is able to reproduce only the dominant orbital-pair correlations.
On average, the values of the mutual information predicted by pCCD differ by approximately an order of magnitude from the DMRG reference values.
As expected, the accuracy of the LCCD and LCCSD corrections strongly depends on the quality of the pCCD reference function, where the overcorrelation introduced by the pCCD model cannot be cured \textit{a posteriori} as observed for the 1D Hubbard model. 
Most importantly, the LCCSD model on top of pCCD breaks down for $U/t \ge 4$, where the eigenvalue spectra of the 2O-RDM contain negative eigenvalues.
For $U/t \ge 4$, the localized pCCD dimer basis represents a suboptimal basis for DMRG and post-pCCD calculations as the single-site entropy features large values approaching the limit $s_i=\ln(4)$.
This break-down of pCCD-LCCSD may also be attributed to the simple response treatment of the $N$-RDMs, where orbital-relaxation effects have not been accounted for.
Nonetheless, we do not expect that accounting for those orbital relaxation effects in the response $N$-RDMs will cure the failure of the pCCD-LCC description in the strong correlation regime as the overcorrelation originates from the pCCD reference function (including the optimization of the on-site basis).
To conclude, the LCCD and LCCSD models can be considered reliable corrections to account for the missing correlation effects on top of a pCCD reference function at least for molecules at their equilibrium geometry or the weak correlation regime.

\section{Acknowledgement}\label{sec:acknowledgement}
A.N. and K.B.~acknowledge financial support from a SONATA BIS grant of the National Science Centre, Poland (no.~2015/18/E/ST4/00584).
\"{O}.L. acknowledges financial support from the Hungarian National Research, Development and Innovation Office 
(K120569) and the Hungarian Quantum Technology National Excellence 
Program (2017-1.2.1-NKP-2017-00001).
The development of the DMRG libraries was supported by the Center for
Scalable and Predictive methods for Excitation and Correlated phenomena (SPEC), which is funded from the Computational Chemical Sciences Program by the U.S. Department of Energy (DOE), at Pacific Northwest National Laboratory.

\normalem

\end{document}